\documentclass[pra,aps,twocolumn,superscriptaddress]{revtex4-1}
\usepackage{graphicx,graphics,xcolor,amsmath,mathtools,
subfigure,bm,bbm,tensor,bbold,amssymb}
\usepackage{hyperref}
\usepackage{cleveref}
\usepackage{stmaryrd}
\usepackage{dcolumn}
\usepackage{float}
\usepackage{calc}
\usepackage{scalerel}

\crefname{equation}{Eq.}{Eqs.}
\Crefname{equation}{Equation}{Equations}
\crefname{figure}{Fig.}{Figs.}
\Crefname{figure}{Figure}{Figures}
\crefname{section}{Sec.}{Secs.}
\crefname{subsection}{Subsec.}{Subsecs.}
\Crefname{section}{Section}{Sections}
\crefname{appendix}{Appendix}{Appendices}
\Crefname{appendix}{Appendix}{Appendices}

\newcommand{\bra}[1]{\ensuremath{ \langle #1  \vert}}
\newcommand{\ket}[1]{\ensuremath{ \vert #1 \rangle}}

\newcommand{\ketbig}[1]{\ensuremath{\big \vert #1\big \rangle}}

\begin{document}

\title{Heralded preparation of  polarization entanglement via quantum scissors}

\author{Dat Thanh Le} \email{thanhdat.le@uq.net.au}
\affiliation{ARC Centre for Engineered Quantum System, School of Mathematics and Physics, University of Queensland, Brisbane, QLD 4072, Australia}
\affiliation{Thang Long Institute of Mathematics and Applied Sciences (TIMAS), Thang Long University, Nghiem Xuan Yem, Hoang Mai, Hanoi 10000, Vietnam}

\author{Warit Asavanant}
\affiliation{Department of Applied Physics, School of Engineering, The University of Tokyo, 7-3-1 Hongo, Bunkyo-ku, Tokyo 113-8656, Japan}

\author{Nguyen Ba An}
\affiliation{Thang Long Institute of Mathematics and Applied Sciences (TIMAS), Thang Long University, Nghiem Xuan Yem, Hoang Mai, Hanoi 10000, Vietnam}
\affiliation{ Institute of Physics, Vietnam Academy of Science and Technology (VAST),
18 Hoang Quoc Viet, Cau Giay, Hanoi 10000 Vietnam }

\begin{abstract}

Quantum entanglement is at the heart of quantum information sciences and quantum technologies. 
In the optical domain, the most common type of quantum entanglement is polarization entanglement, which is 
usually created in a postselection manner involving destructive photon detection and thus hindering further applications which require readily available entanglement resources. In this work, we propose a scheme to prepare multipartite entangled states of polarized photons in a heralded manner, i.e., without postselection.
We exploit the quantum scissors technique to truncate a given continuous-variable entanglement 
into the target entangled states which are of hybrid discrete-continuous or solely discrete types. We consider two implementations of the quantum scissors: one modified from the original quantum scissors [Pegg \textit{et al}., Phys.\ Rev.\ Lett.\ \textbf{81}, 1604 (1998)] using single photons and linear optics and the other designed here using a type-II two-mode squeezer. We clarify the pros and cons of these two implementations as well as discussing practical aspects of the entanglement preparation. 
Our work illustrates an interface between various types of optical entanglement and the proposed quantum scissors techniques could serve as  alternative methods for heralded generation of polarization entanglement.

\end{abstract}

\maketitle

\section{Introduction}

Quantum entanglement plays an essential role in foundations of quantum mechanics \cite{Einstein35} and proves to be an indispensable resource in quantum technologies \cite{NielsenBook00}. Quantum entanglement exists in various forms, among  quantum systems of the same natures or of distinct natures as well as among different degrees of freedom (DoFs) within a single quantum system. Of particular interest is the hybrid entanglement between discrete-variable (DV) and continuous-variable (CV) quantum systems \cite{Jeong14,Morin14,Kwon15,Man17,Rab17,An18,Sychev18,Li18,Huang19,Podoshvedov19,Gouzien20,Le21,Podoshvedov21}, which has been successfully applied in numerous quantum information protocols \cite{Furusawa11Book,Takeda13,Lee13,Kwon13,Takeda15,Andersen15,Ulanov17,Jeannic18,Cavailles18,Guccione20}.

In the optical domain, the most frequently used quantum entanglement is between polarized photons, due to their resilience against decoherence and loss \cite{Kok07,Pan12} and the availability of high-quality polarization-control elements \cite{Kwiat95,KokBook10}. Polarization entanglement was exploited in quantum dense coding \cite{Mattle96}, quantum teleportation \cite{Bouwmeester97}, quantum cryptography \cite{Naik00}, and tests of Bell inequality \cite{Giustina15}. Entangled polarization pairs, called polarization Bell pair (PBP), are routinely produced via  spontaneous parametric downconversion (SPDC), in which a photon in a pumping beam is converted into 
two photons of lower frequencies obeying both
the energy and momentum conservation constraints \cite{Kwiat98}. This process, however, is highly probabilistic and the generated state is dominantly occupied by the unwanted vacuum component \cite{Hnilo05}, which necessitates photon detection to verify entanglement and thus destroys the entangled state itself. 
Such postselection procedure hurdles subsequent quantum information tasks which require an on-demand entanglement resource, such as  quantum error correction \cite{Shor95} and entanglement purification \cite{Bennett96}. 

Therefore, many works have been devoted to the preparation of  polarization entanglement without postselection, e.g., by employing ancilla single photons and linear optics \cite{Zou02,Zou02Third} or a probabilistic controlled-NOT (CNOT) gate \cite{Pittman03}.
In 2010, two groups independently reported heralded generation of a PBP \cite{Barz10,Wagenknecht10}, based on detection of four photons in a three-pair SPDC emission event \cite{Sliwa03}. This method nevertheless suffers from a very low success probability and false detections from a four-pair emission \cite{Kok10Review}.  
Recently, with the development of deterministic, highly pure single-photon sources \cite{Senellart17,Tomm21}, heralded production of a PBP by fusion gate \cite{Bao07,Zhang08}  has been realized \cite{Li21}. The fusion-based method might face difficulty when generalizing to $n$-partite entanglement with $n \geq 3$, which involves nontrivial analyses to find the optimal heralded setup \cite{Gubarev20}.

In this paper, we propose a scheme to prepare polarization entanglement in a heralded fashion, i.e., without relying on postselection. We show that truncating unwanted components in a CV polarization entanglement, which can be supplied by currently accessible resources, gives rise to a hybrid DV-CV or solely DV entangled polarization state of $n$ parties for an arbitrary $n \geq 2$.
We present two implementations for such truncation operation: the first one is a modified version  from the quantum scissors originally proposed in Ref.\ \cite{Pegg98}  using  single photons and linear optics and the second one is proposed here exploiting a type-II two-mode squeezer \cite{KokBook10}.  The technological ingredients  needed for these implementations are well-studied subjects and utilized   in a variety of applications \cite{Ozdemir01,Babichev03,Goyal13,Winnel20,He21,Seshadreesan20,Linares02,Resch02,Zavatta04,Parigi07}. 
Our scheme highlights the intriguing interface between different types of quantum entanglement  and is realizable within the present-day optical technologies.

The outline of this paper is the following. In \cref{sec:Connection} we introduce a particular CV entanglement and show its connection to hybrid DV-CV  and solely DV (i.e., DV-DV) entangled states via truncation of irrelevant terms. We then in \cref{sec:Implementation} propose two distinct methods to implement the desired truncation operation. Next, in \cref{sec:Performance} we analyze the performance for the generation of several types of entanglement using the two truncation techniques. This is followed by discussions on practical perspectives  of the entanglement preparation in \cref{sec:Discussion}. Finally, we conclude the paper in \cref{sec:Conclusion}.  An appendix provides detailed calculations for the results in the main text. 

\section{Truncation of continuous-variable entanglement} \label{sec:Connection}

\begin{figure}[h!]
    \centering
    \includegraphics[width=0.48\textwidth]{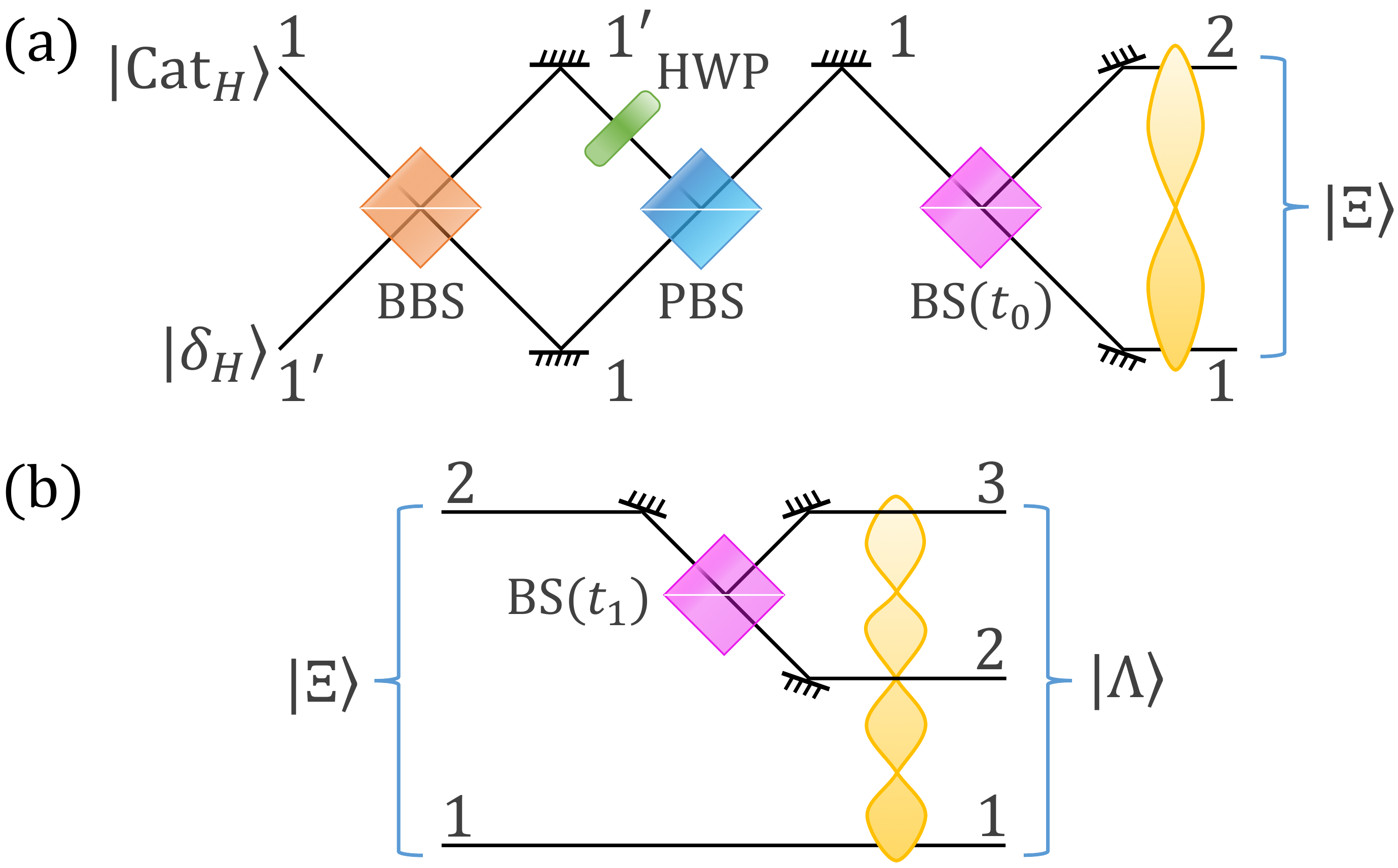}
    \caption{ (a)  Schematic setup to prepare the CV polarization entanglement $\ket{\Xi}_{12}$ in \cref{eq:CVEntanglement2modes}. Here $\ket{\mathrm{Cat}_H} = N_0 (\ket{\delta_H} + e^{i\varphi} \ket{\!-\delta_H}) $ with $N_0 = [2(1+\cos(\varphi) e^{-2\delta^2})]^{1/2} $, BBS is balanced beam splitter, PBS polarizing beam splitter, $\mathrm{BS}(t_0)$ beam splitter with transmissivity $t_0$, and HWP half-wave plate. (b) Splitting of mode 2 in the state $\ket{\Xi}_{12}$ by $\mathrm{BS}(t_1)$ prepares the tripartite CV polarization entanglement $\ket{\Lambda}_{12\dots n}$ with $n=3$ in Eq.\  \eqref{eq:CVEntanglementnmodes}. 
    }
    
    \label{fig:CVEntanglement2modes}
\end{figure}

In what follows, we show that a CV entanglement contains within itself a hybrid DV-CV or a DV-DV entanglement. We consider the CV polarization entanglement \cite{Le21,Bowen02}
\begin{equation}
 \ket{\Xi}_{12} =    N_0 \left (\ket{\alpha_{H}}_{1} \ket{\beta_H}_2 + e^{i \varphi} \ket{\!-\alpha_V}_1 \ket{\!-\beta_V}_2 \right),  \label{eq:CVEntanglement2modes}
\end{equation}
where $N_0 = [2(1 + \cos (\varphi) e^{- (\alpha^2 + \beta^2)})]^{-1/2}$ and $\ket{\gamma_{H/V}} = \sum_{n=0}^{\infty} f_n(\gamma)  \ket{n_{H/V}} $ is a  horizontally or vertically polarized  coherent state of real amplitude $\gamma$ with $f_n(\gamma) = e^{-\gamma^2/2} \gamma^n/\sqrt{n!}$ and $\ket{n_{H/V}}$ a Fock state containing $n$ horizontally or vertically polarized photons. Note that in this paper without loss of generality we consider coherent states of real amplitudes only. For convention the polarization-mode and spatial-mode subscripts are  placed respectively inside and outside of the ket/bra states and from now on, ``mode",  when being used, implies ``spatial mode". Throughout the paper, for brevity we also suppress the vacuum state when expressing ket/bra states as follows: $\ket{\alpha_H,0_V} \equiv \ket{\alpha_H}$, $\ket{0_H,\alpha_V}\equiv \ket{\alpha_V}$, $\ket{n_H,0_V} \equiv \ket{n_H}$, and $\ket{0_H,n_V} \equiv \ket{n_V}$. The vacuum state in some equations will be made visible to perform relevant calculations.

 A schematic setup for the preparation of the state $\ket{\Xi}_{12}$ is shown in \cref{fig:CVEntanglement2modes}a and can be briefly described as follows. The required inputs include  a polarized cat state $\ket{\mathrm{Cat}_H} = N_0 (\ket{\delta_H} + e^{i\varphi} \ket{\!-\delta_H}) $ with $N_0 = [2(1+\cos(\varphi) e^{-2\delta^2})]^{1/2} $ in mode $1$ and a polarized coherent state $\ket{\delta_H}$ in mode $1'$, both of which have the same horizontal polarization. The two states first interact at  a balanced beam splitter (BBS). Since the action of a general beam splitter (BS) with transmissivity $t$ on a pair of coherent states of the same polarization is
\begin{equation}
    \mathrm{BS}_{ab}(t) \ket{\mu}_a\ket{\nu}_b = \ketbig{\mu \sqrt{t} + \nu \sqrt{1-t} }_a \ketbig{\mu \sqrt{1-t} - \nu \sqrt{t}}_b, 
\end{equation}
the two inputs become a NOON-like state \cite{Joo11} $N_0(\ket{\delta\sqrt{2}_H}_1 \ket{0}_{1'} + e^{i\varphi} \ket{0}_1 \ket{\!-\delta\sqrt{2}_H}_{1'})$.
A half-wave plate (HWP) placed at mode $1'$, followed by a polarizing beam splitter (PBS) to merge two modes 1 and $1'$ into one mode 1, changes the the NOON-like state into $N_0(\ket{\delta\sqrt{2}_H}+ e^{i\varphi}  \ket{\!-\delta\sqrt{2}_V})_1$.
A final BS with transmissivity $t_0$ splits such state into two spatial modes, resulting in the state $\ket{\Xi}_{12}$ in \cref{eq:CVEntanglement2modes} with
\begin{equation}
    \alpha = \delta \sqrt{2t_0}, \hspace{1cm} \beta= \delta \sqrt{2(1-t_0)}. \label{eq:AlphaBeta}
\end{equation}

Using Fock-state representation for coherent states, we decompose $\ket{\Xi}_{12}$  as
\begin{eqnarray}
\ket{\Xi}_{12} &=& N_0 f_1 (\alpha)  \left( \ket{1_H}_1 \ket{\beta_H}_2  -  e^{i\varphi} \ket{1_V}_1 \ket{\!-\beta_V}_2 \right)  +  \dots \nonumber \\ \label{eq:Decom1}  \\
&=&  N_0 f_1(\beta)  \left( \ket{\alpha_H}_1 \ket{1_H}_2 - e^{i\varphi} \ket{\!-\alpha_V}_1 \ket{1_V}_2 \right)  +  \dots \nonumber \\ \label{eq:Decom2} \\
&=& N_0 f_1(\alpha) f_1 (\beta)  \left( \ket{1_H}_1 \ket{1_H}_2 + e^{i\varphi} \ket{1_V}_1 \ket{1_V}_2 \right) \nonumber \\
&& +  \dots,  \label{eq:Decom3}
\end{eqnarray}
where in the first equation we hide terms in which mode 1 is not a single-photon state, in the second equation we hide terms in which mode 2 is not a single-photon state, and in the third equation we hide terms in which both modes 1 and 2 are not single-photon states. The visible terms in Eq.\ \eqref{eq:Decom1} or \eqref{eq:Decom2} constitute a hybrid DV-CV entangled state between single photons and coherent states \cite{Le21,Kwon15}, whereas those in Eq.\ \eqref{eq:Decom3} represent a DV-DV entangled state in the form of a PBP.
The decompositions suggest that we can prepare these types of entangled states by performing on mode $1$ and/or mode $2$ of the state $\ket{\Xi}_{12}$ an operation that (i) truncates the non-single-photon components (which include the vacuum state and the more-than-one-photon components), (ii)  preserves coherence of two single-photon states of orthogonal polarizations, and (iii) is carried out in a heralded fashion. We call such an operation the ideal (or perfect) polarized-single-photon quantum scissors and for brevity we abbreviate them as PQS.
The first two requirements for the PQS are obvious, while the last one ensures that the entanglement preparation of interest is heralded given that the input entangled state $\ket{\Xi}_{12}$ is supplied on-demand.  
labeled
Intriguingly, the above observation can be generalized  to the case of an entangled state among $n$ parties for an arbitrary $n$. This is due to the fact that the $n$-partite version of the state $\ket{\Xi}_{12}$, which is of the form
\begin{eqnarray}
\ket{\Lambda}_{12\dots n} &=& M_{n} \big( \ket{\alpha^{(1)}_H}_1 \ket{\alpha^{(2)}_H}_2 \dots \ket{\alpha^{(n)}_H}_n \nonumber \\
&& + e^{i\varphi} \ket{\!-\alpha^{(1)}_V}_1 \ket{\!-\alpha^{(2)}_V}_2 \dots \ket{\!-\alpha^{(n)}_V}_n \big), \quad \label{eq:CVEntanglementnmodes}
\end{eqnarray}
with $M_{n} = [2(1 + \cos (\varphi) \exp( {-\sum_{j=1}^n (\alpha^{(j)})^2} ) ]^{-1/2}$, can be straightforwardly prepared employing $(n - 2)$ BSs to keep splitting $\ket{\Xi}_{12}$. For example, as depicted in \cref{fig:CVEntanglement2modes}b we split mode 2 of the state $\ket{\Xi}_{12}$ by one BS with transmissivity $t_1$ to produce a tripartite CV polarization entangled state
\begin{equation}
    N_0 \big( \ket{\alpha^{(1)}_H}_1 \ket{\alpha^{(2)}_H}_2 \ket{\alpha^{(3)}_H}_3 + e^{i\varphi} \ket{\!-\alpha^{(1)}_V}_1 \ket{\!-\alpha^{(2)}_V}_2 \ket{\!-\alpha^{(3)}_V}_3 \big),
\end{equation}
where $\alpha^{(1)} = \delta\sqrt{2t_0}$, $\alpha^{(2)} = \delta\sqrt{2(1-t_0)t_1} $, and $\alpha^{(3)} = \delta\sqrt{2(1-t_0)(1-t_1)} $. Generalization to a higher number of parties is  straightforward.
Given the state $\ket{\Lambda}_{12\dots n}$, we perform the PQS on $j$ parties of it, say from party $1$ to party $j$, to obtain an $n$-partite hybrid DV-CV entangled state of the form
\begin{eqnarray}
\ket{\Omega}_{12\dots n} \! &=& \! \frac{1}{\sqrt{2}} (\ket{1_H}_1 \! \dots \! \ket{1_H}_{j} \ket{\alpha^{(j+1)}_H}_{j+1} \! \dots \! \ket{\alpha^{(n)}_H}_{n} \nonumber \\
&& + e^{i\varphi} \ket{1_V}_1 \! \dots \! \ket{1_V}_{j} \ket{\!-\alpha^{(j+1)}_V}_{j+1} \! \dots \! \ket{\!-\alpha^{(n)}_V}_{n} ). \nonumber \\ \label{eq:NPartiteHy}
\end{eqnarray}
When $j=n$ the above state becomes the $n$-partite DV GHZ polarization entanglement. 

\section{Implementations of the polarized-single-photon quantum scissors} \label{sec:Implementation}

 In this section we present two implementations of the PQS that truncates the CV polarization entanglements in Eqs.\ \eqref{eq:CVEntanglement2modes} and \eqref{eq:CVEntanglementnmodes}  into hybrid DV-CV and   DV-DV entangled states. We show that the two implementations actually realize a nonideal PQS in the sense that they perfectly satisfy the requirements (ii) (preserving coherence of two orthogonally polarized single photons) and (iii) (succeeding in a heralded way) but partially meet the 
requirement (i) (see the requirements for the ideal PQS in \cref{sec:Connection}). That is, 
 they do not completely truncate all the non-single-photon components but retain the vacuum and a two-photon state, resulting in an unwanted imperfection. This imperfection, however, can be made arbitrarily small by suitably adjusting relevant parameters.

\subsection{Using single photons and linear optics} \label{subsec:FirstImplementation}

The first implementation of the desired PQS, denoted as PQS1, involves the original quantum scissors  proposed in Refs.\ \cite{Pegg98,Winnel20} which truncate the more-than-one-photon components and amplify the one-photon component compared to the vacuum one in a quantum state 
\begin{equation}
    \ket{\psi} = \sum_{n=0}^{\infty} c_n \ket{n}, 
\end{equation}
where $\ket{n}$ is a Fock state containing $n$ photons and $\sum_{n=0}^{\infty} |c_n|^2=1$. We refer to such scissors as QS to distinguish from the PQS.
The QS deals with photons having only one polarization, whereas the PQS is supposed to work with inputs having two orthogonal polarizations. Therefore, the QS is not readily applicable for extracting the desired entanglements out of the CV entanglement in \cref{eq:CVEntanglement2modes} or \eqref{eq:CVEntanglementnmodes}. In the following, we first present some details of the QS and then show how to construct the PQS1 from it.

\begin{figure}[ht!]
    \centering
    \includegraphics[width=0.48 \textwidth]{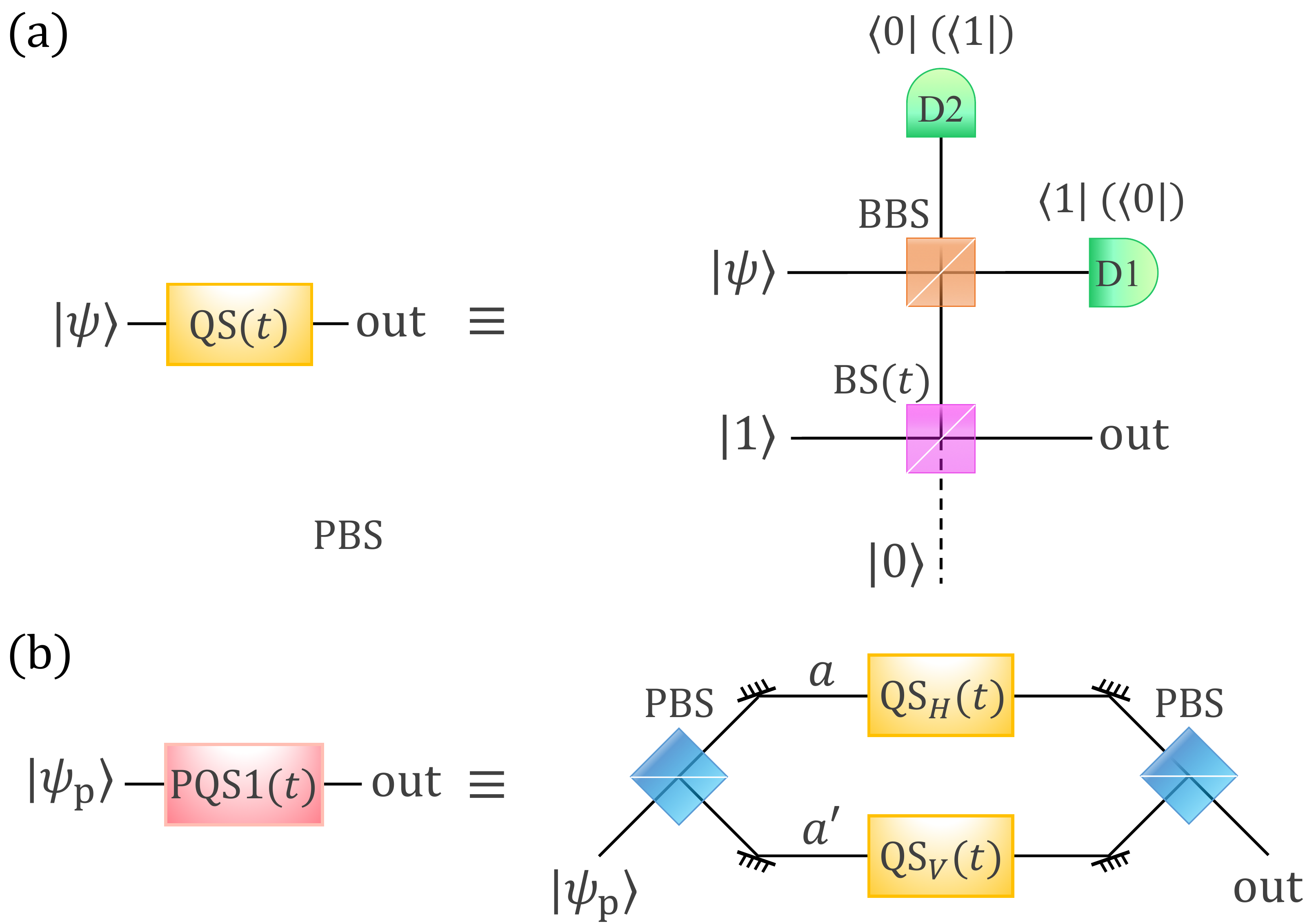}
    
    \caption{(a) Black-box representation with the label $\mathrm{QS}(t)$ and physical setup  of the quantum scissors proposed in Refs.\ \cite{Pegg98,Winnel20}. Here $t$ denotes the transmissitivity of the BS, which characterizes the performance of the scissors.  The input $\ket{\psi} = \sum_{n=0}^{\infty} c_n \ket{n}$, going through the $\mathrm{QS}(t)$, is truncated into the (unnormalized) output state $  \sqrt{1-t} c_0 \ket{0} \pm \sqrt{t} c_1 \ket{1}$, conditioned by detection of a single photon  at detector D1 and no photons  at detector D2 or no photons at detector D1 and a single photon at detector D2. Choosing $t$ properly this output  can be made very close to a single-photon state. (b) Linear-optics-based implementation of a nonideal PQS with the black-box labeled as $\mathrm{PQS}1(t)$, comprising two modules of the quantum scissors in panel (a). The $\mathrm{PQS1}(t)$ truncates  a general polarized state $\ket{\psi_{\mathrm{p}}} = \sum_{n,m=0}^{\infty} c_{nm} \ket{n_H,m_V}$ into an output state that is very close to the (unnormalized) state $\ket{\psi_{\mathrm{p}}^{(1)}} = c_{10}\ket{1_H,0_V} + c_{01}\ket{1_V,0_H}$.  The box with the label $\mathrm{QS}_H(t)$ ($\mathrm{QS}_V(t)$) represents the physical setup in panel (a) with the ancilla single photon being horizontally (vertically) polarized.} 
    
    \label{fig:QuantumScissors}
\end{figure}

The black-box representation of the QS and its physical implementation are depicted in \cref{fig:QuantumScissors}a.  
Concretely, following the schematic setup in  \cref{fig:QuantumScissors}a the QS transforms the state $\ket{\psi}$ as
 \begin{equation}
    \ket{\psi} = \sum_{n=0}^{\infty} c_n \ket{n} \xrightarrow{\mathrm{QS}(t)} \sqrt{1-t} c_0 \ket{0} \pm   \sqrt{t} c_1 \ket{1}   , \label{eq:qSDiagram}
\end{equation}
where the output state is unnormalized and the relative sign ``$+$" \mbox{(``$-$")} corresponds to detection of a single photon (vacuum) at detector D1 and vacuum (a single photon) at detector D2. 
We choose the BS transmissitivity $t$ such that $ \sqrt{1-t}c_0$ is much smaller than $\sqrt{t}c_1 $, making the output state  close enough to a single photon. The setup in \cref{fig:QuantumScissors}a resembles an error-corrected quantum teleportation circuit, of which the state $\ket{\psi}$ serves as the input, the quantum channel is a   single-rail entangled state created by splitting the single-photon input via the $\mathrm{BS}(t)$, and the Bell measurement is performed using the BBS and two detectors D1 and D2. The output state is corrected, compared to the input one $\ket{\psi}$, in the sense that more-than-one-photon components are erased, recovering $\ket{\psi}$ back to the single-rail basis. The heralding probability and the fidelity of the QS technique are respectively
\begin{eqnarray}
P_{\mathrm{QS}} &=& (1-t) |c_0|^2 + t |c_1|^2 , \\
F_{\mathrm{QS}} &=& \frac{t |c_1|^2}{(1-t) |c_0|^2 + t |c_1|^2}. 
\end{eqnarray}
The probability $P_{\mathrm{QS}}$ here is nothing but the inverse square of the normalization factor of the output state in \cref{eq:qSDiagram}. 

The above result also holds when the to-be-truncated mode is entangled with others. To verity this, let us consider a quantum state comprising spatially distinguishable modes $a,b,c,\dots$ in the form
\begin{equation}
    \ket{\Psi}_{abc\dots} = \sum_{n=0}^{\infty} c_n \ket{n}_a \ket{\phi_n}_{bc\dots},
\end{equation}
where $a$ is the to-be-truncated mode and $\ket{\phi_n}_{bc\dots}$ is a (normalized) joint quantum state of modes $b,c\dots$. We can check that the QS  in \cref{fig:QuantumScissors}a  applying  to only mode $a$ reduces $\ket{\Psi}_{abc\dots}$ in the following way
 \begin{eqnarray}
 && \ket{\Psi}_{abc\dots} = \sum_{n=0}^{\infty} c_n \ket{n}_a \ket{\phi_n}_{bc\dots} \nonumber \\
 &&\xrightarrow[\text{on\,mode}\,a]{\mathrm{QS}(t)}  
 \sqrt{1-t} c_0 \ket{0}_a \ket{\phi_0}_{bc\dots} + \sqrt{t} c_1 \ket{1}_a \ket{\phi_1}_{bc\dots} . \qquad \label{eq:qSDiagram2Modes}
 \end{eqnarray}
 This is just the same as the result in \cref{eq:qSDiagram} if we formally make the replacement $c_n \to c_n \ket{\phi_n}_{bc\dots} $. We also note that different from \cref{eq:qSDiagram} the relative sign in the output state of Eq.\ \eqref{eq:qSDiagram2Modes} has been chosen to be ``$+$" for definiteness.

We now turn to the design of the PQS1. We consider a single-spatial-mode polarized quantum state of a general form
\begin{equation}
\ket{\psi_{\mathrm{p}}} = \sum_{n,m=0}^{\infty} c_{nm} \ket{n_H,m_V}, \label{eq:generalPolarizedQuantumState}
\end{equation}
where $\ket{n_H,m_V}$ describes a quantum state of the single spatial mode of interest having $n$ horizontally polarized photons and $m$ vertically polarized photons and $\sum_{n,m=0}^{\infty}|c_{nm}|^2=1$. 
The concerned PQS1 should operate in such a way that it truncates the state $\ket{\psi_{\mathrm{p}}}$  into (exactly or very close to) the target output state
\begin{equation}
  \ket{\psi_{\mathrm{p}}^{(1)}} =  c_{10}\ket{1_H,0_V} + c_{01} \ket{0_H,1_V} \equiv c_{10}\ket{1_H} + c_{01} \ket{1_V}. \label{eq:PolIdealState}
\end{equation}
For this purpose, we arrange a setup including two modules of the QS, $\mathrm{QS}_H(t)$ and $\mathrm{QS}_V(t)$, as sketched in \cref{fig:QuantumScissors}b, where $\mathrm{QS}_{H/V}(t)$ operates as  $\mathrm{QS}(t)$ in \cref{fig:QuantumScissors}a when the input photon is horizontally or vertically polarized. According to \cref{fig:QuantumScissors}b, we first separate spatially two polarizations of the single-spatial-mode state  $\ket{\psi_{\mathrm{p}}}$ by a PBS to obtain a two-spatial-mode state
\begin{equation}
    \sum_{n,m=0}^{\infty} c_{nm} \ket{n_H}_a \ket{m_V}_{a'} \equiv \sum_{n=0}^{\infty} \ket{n_H}_a  \sum_{m=0}^{\infty} c_{nm} \ket{m_V}_{a'}.  \label{eq:Psip2mode}
\end{equation}
Following the transformation in Eq.\ \eqref{eq:qSDiagram2Modes}, the module $\mathrm{QS}_H (t)$ acting on mode $a$ truncates the state in \cref{eq:Psip2mode} into
\begin{eqnarray}
 &&  \! \sqrt{1-t} \ket{0_H}_{a}  \sum_{m=0}^{\infty}c_{0m}  \ket{m_V}_{a'}  \! + \! \sqrt{t} \ket{1_H}_{a}  \sum_{m=0}^{\infty}c_{1m} \ket{m_V}_{a'}  \nonumber \\
    &\equiv & \! \sum_{m=0}^{\infty} \ket{m_V}_{a'} ( \sqrt{1-t} c_{0m} \ket{0_H}   \! +\! \sqrt{t} c_{1m} \ket{1_H} )_a . \label{eq:Psip2modeQSH}
\end{eqnarray}
At the same time, the module $\mathrm{QS}_V (t)$ on mode $a'$, also according to Eq.\ \eqref{eq:qSDiagram2Modes}, shortens the state in Eq.\ \eqref{eq:Psip2modeQSH} into
\begin{eqnarray}
 && \sqrt{1-t} \ket{0_V}_{a'} ( \sqrt{1-t} c_{00} \ket{0_H} + \sqrt{t} c_{10} \ket{1_H}     )_a
 \nonumber \\
  && +  \sqrt{t} \ket{1_V}_{a'} (\sqrt{1-t} c_{01} \ket{0_H} + \sqrt{t} c_{11} \ket{1_H}    )_a . \qquad
\end{eqnarray}
This truncated two-spatial-mode state going through another PBS and after rearrangements is recast to a single-spatial-mode state 
\begin{equation}
   \sqrt{(1-t)t} \ket{\psi_{\mathrm{p}}^{(1)}}_a    + (1-t) c_{00} \ket{0_H,0_V}_{a} + t c_{11} \ket{1_H,1_V}_a. \label{eq:OutputPolqS}
\end{equation}

We extend the above result to the case when the mode to be truncated is in entanglement with others. That is, instead of the state $\ket{\psi_{\mathrm{p}}}$ in \cref{eq:generalPolarizedQuantumState} we consider
\begin{equation}
    \ket{\Psi_{\mathrm{p}}}_{abc\dots} = \sum_{n,m=0}^{\infty} c_{nm} \ket{n_H,m_V}_a \ket{\phi_{nm}}_{bc\dots}, \label{eq:GeneralEntangledState}
\end{equation}
where $a$ is the to-be-truncated mode and $\ket{\phi_{nm}}_{bc\dots}$ is a (normalized) joint quantum state of modes $b,c\dots$. We aim to get the ideal output state after truncation as
\begin{eqnarray}
\ket{\Psi_{\mathrm{p}}^{(1)}}_{abc\dots}&=& c_{10} \ket{1_H, 0_V}_a \ket{\phi_{10}}_{bc\dots}  \!+\!  c_{01} \ket{0_H, 1_V}_a \ket{\phi_{01}}_{bc\dots} \nonumber \\
&\equiv & c_{10} \ket{1_H}_a \ket{\phi_{10}}_{bc\dots} \! + \! c_{01} \ket{1_V}_a \ket{\phi_{01}}_{bc\dots}. \label{eq:PolIdealEntangledState}
\end{eqnarray}
 Similar to the output in Eq.\ \eqref{eq:OutputPolqS}, application of the PQS1 on mode $a$ in the state $\ket{\Psi_{\mathrm{p}}}_{abc\dots}$ gives
\begin{eqnarray}
 &&   \sqrt{(1\!-\!t)t} \ket{\Psi_{\mathrm{p}}^{(1)}}_{abc\dots}  + (1-t) c_{00} \ket{0_H,\! 0_V}_a \ket{\phi_{00}}_{bc\dots} \nonumber \\
 && + t c_{11} \ket{1_H,1_V}_a \ket{\phi_{11}}_{bc\dots} .  \label{eq:OutputPQS1PsiP}
\end{eqnarray}
The success probability and the corresponding fidelity between this output state and the ideal one $\ket{\Psi_{\mathrm{p}}^{(1)}}_{abc\dots}$  are 
\begin{eqnarray}
 P_{\mathrm{PQS1}} &=& (1-t)t (|c_{10}|^2 \! + \! |c_{01}|^2) \!  + \! (1-t)^2 |c_{00}|^2  \! + \! t^2 |c_{11}|^2, \nonumber \\  \label{eq:ProbPQS1PsiP} \\
 F_{\mathrm{PQS1}} &=& \frac{(1-t)t (|c_{10}|^2 + |c_{01}|^2)}{(1-t)t (|c_{10}|^2 \!+ \! |c_{01}|^2)  \! + \! (1-t)^2 |c_{00}|^2 \! + \! t^2 |c_{11}|^2}. \nonumber \\ \label{eq:FidePQS1PsiP}
\end{eqnarray}
As will be shown later, for our initial input entanglement of interest, i.e., the state $\ket{\Psi_{\mathrm{p}}}_{abc\dots}$ in  \cref{eq:GeneralEntangledState}, $c_{11}=0$, so that by choosing $t \to 1$ the fidelity $F_{\mathrm{PQS1}}$ is approaching 1 but the  success probability $P_{\mathrm{PQS1}}$ is turning out to be very low.

\subsection{Using type-II two-mode squeezer} \label{subsec:SecondImplementation}

\begin{figure}[ht!]
    \centering
    \includegraphics[width=0.48\textwidth]{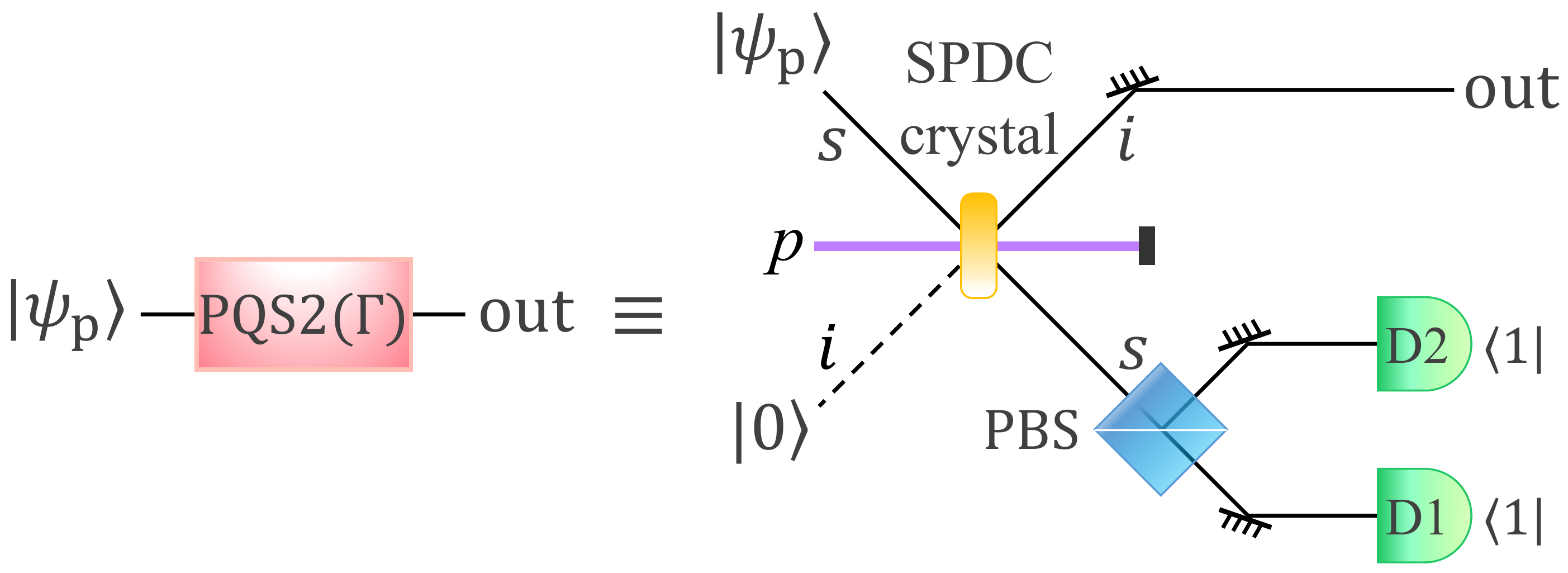}
    \caption{Two-mode-squeezer-based  implementation of a nonideal PQS with the black box labeled as $\mathrm{PQS}2(\Gamma)$ using a type-II two-mode squeezer represented by a  SPDC crystal \cite{Linares02}. Here $\Gamma$ is the squeezing parameter characterizing the performance of the scissors.  The input state $\ket{\psi_{\mathrm{p}}} = \sum_{n,m=0}^{\infty} c_{nm} \ket{n_H,m_V}$ is injected into the signal ($s$) mode of the squeezer while the idle ($i$) mode is in the vacuum. Co-detection of single photons at both detectors D1 and D2 heralds the truncated output at the idle mode which is very close to the desired state $\ket{\psi_{\mathrm{p}}^{(1)}} = c_{10}\ket{1_H,0_V} + c_{01}\ket{1_V,0_H}$.
    A pump, denoted by \textit{p}, stimulates squeezing in the SPDC crystal.}
    \label{fig:DatQuantumScissors}
\end{figure}

Here we propose a different PQS implementation, denoted as PQS2, employing a type-II two-mode squeezer
\begin{equation}
    \hat S_{si}  = \exp (\xi  \hat K_{si}^\dag - \xi^*   \hat K_{si} ), \label{eq:Squeezer}
\end{equation}
where $s$ and $i$ respectively denote the signal and idle modes, $\xi$ is proportional to the coupling constant $\chi^{(2)}$ and the intensity of the classical pump \cite{GerryBook04}, and 
\begin{equation}
    \hat K_{si} = \hat a_{s,H} \hat a_{i,V} + \hat a_{s,V} \hat a_{i,H}. \label{eq:K+si}
\end{equation}
This squeezer is typically realized by a SPDC crystal \cite{KokBook10} and commonly used  in laboratories. 
 The  setup to realize the PQS2 for truncating unwanted components in the state $\ket{\psi_{\mathrm{p}}}$ in \cref{eq:generalPolarizedQuantumState} is depicted in \cref{fig:DatQuantumScissors} comprising two main steps: (step 1) injecting $\ket{\psi_{\mathrm{p}}}$ to mode $s$ of the squeezer while leaving mode $i$ in the vacuum, which is mathematically equivalent to
acting $\hat S_{si}$ on $\ket{\psi_{\mathrm{p}}}_s \ket{0}_i$, and (step 2) detecting two photons in mode $s$, one horizontally polarized and the other vertically polarized, which heralds the truncated output state in mode $i$ that is expected to be exactly or very close to the desired state $ \ket{\psi_\mathrm{p}^{(1)}}$ in \cref{eq:PolIdealState}. 

To get intuition on how this PQS2 implementation works, let us consider a scenario in which we inject $ \ket{\psi_\mathrm{p}^{(1)}} = c_{10} \ket{1_H} + c_{01} \ket{1_V} $ in \cref{eq:PolIdealState} to mode $s$ of the squeezer $\hat S_{si}$ and leave mode $i$ in the vacuum. We approximate $\hat S_{si}$ to the first order of $|\xi|$, which is typically of order $10^{-2}$ \cite{KokBook10}, as $ \hat S_{si} \simeq 1 + (\xi \hat K^\dag_{si} - \xi^* \hat K_{si} )$. We find that 
\begin{eqnarray}
 \hat S_{si} \ket{\psi_{\mathrm{p}}^{(1)}}_s \ket{0}_i  &\simeq &   \ket{\psi_{\mathrm{p}}^{(1)}}_s \ket{0}_i  +  + \xi \ket{1_H,1_V}_s \ket{\psi_{\mathrm{p}}^{(1)}}_i \nonumber \\
 && + \xi \sqrt{2} (c_{10}\ket{2_H}_s \ket{1_V}_i  +  c_{01}\ket{2_V}_s \ket{1_H}_i ). \nonumber \\ \label{eq:SqueezerActingOnPsi1}
\end{eqnarray}
We then detect in mode $s$ single photons of both horizontal and vertical polarizations, i.e., we perform  measurement with the projector 
\begin{equation}
  \hat \Pi_s = \ket{1_H,1_V}_s \bra{1_H,1_V} \label{eq:MeasurementOnModes}.
\end{equation}
The projected state in mode $i$ will be $\ket{\psi_{\mathrm{p}}^{(1)}}$, implying that the initial input state of mode $s$ is completely transferred to mode $i$. 
We repeat the same procedure for different inputs in mode $s$, including the vacuum state $\ket{0}$ and $\ket{n_H,m_V}$ with $n,m \geq 1$, and observe that
\begin{eqnarray}
\hat S_{si} \ket{0}_s \ket{0}_i &\simeq& \ket{0}_s \ket{0}_i  + \xi (\ket{1_H}_s \ket{1_V}_i \! + \! \ket{1_V}_s \ket{1_H}_i), \nonumber \\ \label{eq:SqueezerActingOnVacuum} \\
 \hat S_{si} \ket{n_H,m_V}_s \ket{0}_i &\simeq& \ket{n_H,m_V}_s \ket{0}_i  \nonumber \\
 && + \xi \sqrt{n\!+\!1}\ket{(n\!+\!1)_H,m_V}_s\ket{1_V}_i  \nonumber \\
 && + \xi \sqrt{m\!+\!1}\ket{n_H,(m\!+\!1)_V}_s \ket{1_H}_i, \qquad \label{eq:SqueezerActingOnnm}
\end{eqnarray}
of which the probability of finding $\ket{1_H,1_V}_s$ in mode $s$ is non-zero only when $n=m=1$. Noticing that the state $\ket{\psi_{\mathrm{p}}}$ in \cref{eq:generalPolarizedQuantumState} is a superposition of $\ket{\psi_{\mathrm{p}}^{(1)}}$, $\ket{0}$, and $\ket{n_H,m_V}$ with $n,m\geq 1$,  the results in Eqs.\ \eqref{eq:SqueezerActingOnPsi1}, \eqref{eq:SqueezerActingOnVacuum}, and \eqref{eq:SqueezerActingOnnm} thus suggest that
if the input state in mode $s$ is $\ket{\psi_{\mathrm{p}}}$, 
application of $\hat S_{si}$ on the input $\ket{\psi_{\mathrm{p}}}_s \ket{0}_i$ combining with the subsequent measurement $\hat \Pi_s$ on mode $s$ will yield in mode $i$ an output state consisting of
$\ketbig{\psi_{\mathrm{p}}^{(1)}}$ and $\ket{0}$. When the contribution from the vacuum is negligible, this can realize the map $\ket{\psi_{\mathrm{p}}} \to \ket{\psi_{\mathrm{p}}^{(1)}}$, i.e., the PQS2, with a high fidelity. We also note that an unwanted term from second-order squeezing can appear in the output, as will be shown below by exact calculations.   

We facilitate exact calculations related to the squeezer by using the following formula \cite{Simon00}
\begin{eqnarray}
  \hat S_{si} \ket{n_H,m_V}_s \ket{0}_i &=& K_{n+m}  \sum_{k,l=0}^{\infty} (-i\Gamma)^{k+l} (C_{n+k}^n C_{m+l}^m)^{\tfrac{1}{2}} \quad \nonumber \\
  && \times \ket{(n\!+\!k)_H,(m\!+\!l)_V}_s \ket{l_H,k_V}_i, \label{eq:UsefulFormula}
\end{eqnarray}
where $C^k_n = n!/[(n - k)!k!]$, $\Gamma = \tanh(i \xi ) $ is the characteristic squeezing parameter, and $K_n = (1-|\Gamma|^2)^{(n+2)/2}$. Acting $\hat S_{si}$ on $\ket{\psi_{\mathrm{p}}}_s\ket{0}_i$ (step 1) then produces
\begin{eqnarray}
 \hat S_{si} \ket{\psi_{\mathrm{p}}}_s \ket{0}_i \! &=& \!  \sum_{n,m=0}^{\infty}\! \! c_{nm} K_{n+m}  \sum_{k,l=0}^{\infty}\! (-i\Gamma)^{k+l} (C_{n+k}^n C_{m+l}^m)^{\tfrac{1}{2}} \nonumber \\
  && \! \times \ket{(n+k)_H,(m+l)_V}_s \ket{l_H,k_V}_i. \label{eq:ExactSqueezing}
\end{eqnarray}
Performance of the measurement $\hat \Pi_s$ in \cref{eq:MeasurementOnModes} on mode $s$ of this state (step 2) projects mode $i$ onto
 \begin{equation}
   K_1 (-i\Gamma) \ket{\psi_{\mathrm{p}}^{(1)}}  + c_{11}K_2 \ket{0} + c_{00}K_0 (-i \Gamma)^2 \ket{1_H,1_V}, \label{eq:OutputDatQS}
 \end{equation}
 where the last term, which is an unwanted two-photon state, results from second-order squeezing.

We replace the single-mode input $\ket{\psi_p}$ by the state $\ket{\Psi_{\mathrm{p}}}_{abc\dots}$ in \cref{eq:GeneralEntangledState} and apply the PQS2 above to only mode $a$. 
Analogous to the result in Eq.\ \eqref{eq:OutputDatQS} we obtain the following output
\begin{eqnarray}
  && K_1 (-i\Gamma) \ket{\Psi_{\mathrm{p}}^{(1)}}_{abc\dots} + c_{11}K_2 \ket{0}_a \ket{\phi_{11}}_{bc\dots} \nonumber \\
  &&+ c_{00}K_0 (-i \Gamma)^2 \ket{1_H,1_V}_a \ket{\phi_{00}}_{bc\dots}, \label{eq:OutputPQS2PsiP}
 \end{eqnarray}
 where $\ket{\Psi_{\mathrm{p}}^{(1)}}_{abc\dots}$, given in Eq.\ \eqref{eq:PolIdealEntangledState}, is the ideal truncated state. 
 The success probability and the fidelity of the state preparation for this case are
\begin{eqnarray}
 P_{\mathrm{PQS2}} &=&  (|c_{10}|^2\!+\!|c_{01}|^2) K_1^2 |\Gamma|^2 \! + \! |c_{11}|K^2_2 \! + \! |c_{00}|^2 K_0^2 |\Gamma|^4, \nonumber \\
 \\ \label{eq:ProbPQS2PsiP}
 F_{\mathrm{PQS2}} &=& \frac{(|c_{10}|^2\!+\!|c_{01}|^2) K_1^2 |\Gamma|^2  }{(|c_{10}|^2\!+\!|c_{01}|^2) K_1^2 |\Gamma|^2 \! + \! |c_{11}|K^2_2 \! + \! |c_{00}|^2 K_0^2 |\Gamma|^4}. \nonumber \\ \label{eq:FidePQS2PsiP}
\end{eqnarray}
As typically $ |\xi| \sim 10^{-2} \ll 1 $ \cite{KokBook10}, one finds $|\Gamma| \sim 10^{-2} $ and $K_0 \approx K_1 \approx K_2 \approx 1$. This combining with the fact that $c_{11}=0$ for our particular input entanglement, i.e., the state $\ket{\Psi_{\mathrm{p}}}_{abc\dots}$ in \cref{eq:GeneralEntangledState}, hints that the fidelity $F_{\mathrm{PQS2}}$ should be very high. 

\section{Performance analysis of the entanglement preparation} \label{sec:Performance}

\begin{figure}[ht!]
    \centering
    \includegraphics[width=0.48 \textwidth]{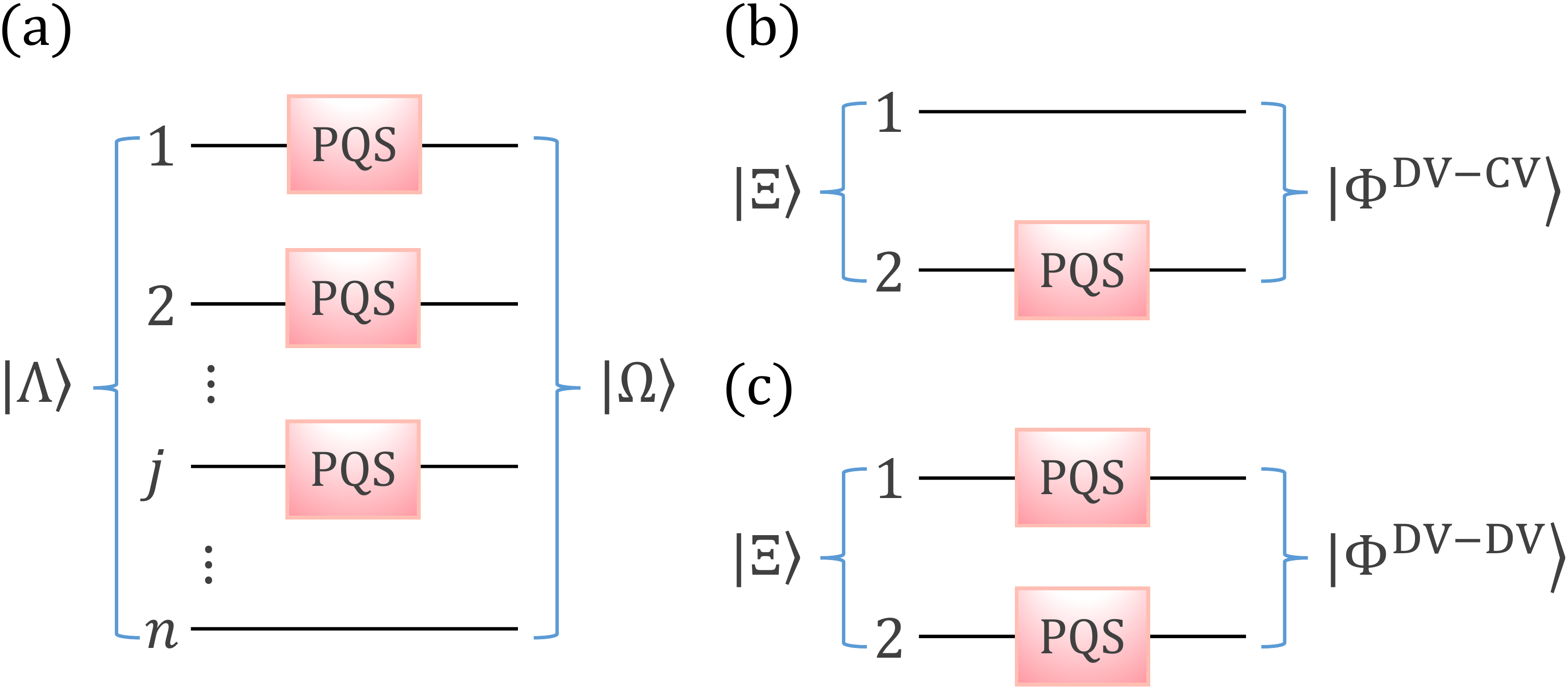}
    \caption{(a) Applying the PQS on $j$ modes of the CV entanglement $\ket{\Lambda}_{12\dots n}$ in Eq.\ \eqref{eq:CVEntanglementnmodes} produces the entanglement $\ket{\Omega}_{12\dots n}$ in Eq.\ \eqref{eq:NPartiteHy}. For $n=2$ and $j=1$, panel (a) reduces to panel (b) which shows the preparation of the hybrid entanglement $\ket{\Phi^{\mathrm{DV\text{-}CV}}}$ in \cref{eq:DVCV} from the CV entanglement $\ket{\Xi}_{12}$ in \cref{eq:CVEntanglement2modes}. For $n=2$ and $j=2$, panel (a) reduces to panel (c) which shows the preparation of the DV PBP $\ket{\Phi^{\mathrm{DV\text{-}DV}}}$ in \cref{eq:DVDV} from the CV entanglement $\ket{\Xi}_{12}$. Here the PQS can be realized by either the PQS1 or the PQS2. }
    \label{fig:EntanglementPrepDiagram}
\end{figure}

At this point, we are ready to exploit the two PQS implementations, the PQS1 and the PQS2, developed in the previous section to prepare the target entangled states. In \cref{fig:EntanglementPrepDiagram}a, we show the generation of the $n$-partite polarization entangled state $\ket{\Omega}_{12\dots n}$ in Eq.\ \eqref{eq:NPartiteHy} using PQSs to truncate $j$ modes of the input CV entanglement $\ket{\Lambda}_{12\dots n}$ in Eq.\ \eqref{eq:CVEntanglementnmodes}. In what follows, we consider two specific cases with $(n,j)=(2,1)$ and $(n,j)=(2,2)$, which correspond respectively to the preparations of the hybrid DV-CV entangled state 
\begin{equation}
    \ket{\Phi^{\mathrm{DV\text{-}CV}}} =   \frac{1}{\sqrt{2}} (\ket{1_H} \ket{\alpha_H}  + e^{i\varphi} \ket{1_V} \ket{\!-\alpha_V}  ),  \label{eq:DVCV}
\end{equation}
and the DV-DV PBP 
\begin{equation}
    \ket{\Phi^{\mathrm{DV\text{-}DV}}} = \frac{1}{\sqrt{2}} (\ket{1_H}\ket{1_H} + e^{i\varphi} \ket{1_V}\ket{1_V} ), \label{eq:DVDV}
\end{equation}
via truncating the input entanglement $\ket{\Xi}_{12}$ in \cref{eq:CVEntanglement2modes}.
The schematic diagrams for the entanglement preparations of interest are shown in Figs. \ref{fig:EntanglementPrepDiagram}b and \ref{fig:EntanglementPrepDiagram}c. 
We compute the success probability and the fidelity when preparing these entangled states by means of both the PQS1 and the PQS2.

\subsection{Hybrid DV-CV entangled state}

We reexpress the state $\ket{\Xi}_{12}$ in \cref{eq:CVEntanglement2modes} in the form of the state $\ket{\Psi_{\mathrm{p}}}_{abc\dots}$ in \cref{eq:GeneralEntangledState}
\begin{equation}
 \ket{\Xi}_{12} = \sum_{n,m=0}^{\infty} c_{nm} \ket{n_H,m_V}_2 \ket{\phi_{nm}}_1, \label{eq:XiRewritten}
\end{equation}
where mode $2$ is to be truncated and
\begin{equation}
 \left\{   \begin{array}{rlrl}
   c_{10} = & N_0 f_1(\beta), &  \ket{\phi_{10}}  = & \ket{\alpha_H}, \\
      c_{01} = & N_0 e^{i\varphi} f_1 (-\beta), & \ket{\phi_{01}}  = & \ket{\!-\alpha_V}, \\
      c_{00}  = & N_0 L_\alpha^{-1} f_0 (\beta), &  \ket{\phi_{00}}   = & L_\alpha (\ket{\alpha_H} \! + \! e^{i\varphi} \ket{\!-\alpha_V}), \\
      c_{11} = &0, & \ket{\phi_{11}} =& \varnothing, \\
     \vdots   & & \vdots  & 
\end{array} \right.
\end{equation}
 with $L_\alpha = [2(1+\cos(\varphi) e^{-\alpha^2})]^{-1/2}$. Using the results in \cref{subsec:FirstImplementation} [i.e., Eqs.\ \eqref{eq:OutputPQS1PsiP} to \eqref{eq:FidePQS1PsiP}] application of the PQS1 on mode 2 of the state $\ket{\Xi}_{12}$ written in the form of \cref{eq:XiRewritten} gives the (unnormalized) output
\begin{eqnarray}
 &&  \sqrt{(1-t)t} N_0 f_1(\beta) (\ket{1_H}_2 \ket{\alpha_H}_1 - e^{i\varphi} \ket{1_V}_2 \ket{\!-\alpha_V}_1 ) \nonumber \\
 && + (1-t) N_0 L_\alpha^{-1} f_0 (\beta) \ket{0}_2 L_\alpha(\ket{\alpha_H} + e^{i\varphi} \ket{\!-\alpha_V})_1. \qquad \label{eq:OutputHybridPQS1}
\end{eqnarray}
The heralding probability and the fidelity of  the prepared state to the hybrid DV-CV  entanglement $( \ket{1_H}_2 \ket{\alpha_H}_1 - e^{i\varphi} \ket{1_V}_2 \ket{\!-\alpha_V}_1 )/\sqrt{2}$, which is up to a local unitary transformation equivalent to the state $\ket{\Phi^{\mathrm{DV\text{-}CV}}}$ in \cref{eq:DVCV}, are
\begin{eqnarray}
 P^{\mathrm{DV\text{-}CV}}_{\mathrm{PQS1}} &=& 2(1-t)t N_0^2 f_1^2(\beta) + (1-t)^2 N_0^2 L_\alpha^{-2} f_0^2(\beta), \nonumber \\ \label{eq:ProbHy1} \\
 F^{\mathrm{DV\text{-}CV}}_{\mathrm{PQS1}} &=& \frac{2t f_1^2(\beta)}{2t f_1^2(\beta) + (1-t) L_\alpha^{-2} f_0^2(\beta)}. \label{eq:FideHy1}
\end{eqnarray}
The unwanted term containing the vacuum in mode 2 of the output state in Eq.\ \eqref{eq:OutputHybridPQS1} results in an additional contribution to the heralding probability $P^{\mathrm{DV\text{-}CV}}_{\mathrm{PQS1}}$ [i.e., the second term in the RHS of Eq.\ \eqref{eq:ProbHy1}] but induces a less-than-one fidelity $F^{\mathrm{DV\text{-}CV}}_{\mathrm{PQS1}}$ [i.e.,  the presence of a second term in the denominator of the RHS of Eq.\ \eqref{eq:FideHy1}]. Notably, as $\alpha$ and $\beta$ are defined via $\delta$ and $t_0$ as in \cref{eq:AlphaBeta} and $N_0$ is defined via $\delta$ and $\varphi$, $P^{\mathrm{DV\text{-}CV}}_{\mathrm{PQS1}}$ and $F^{\mathrm{DV\text{-}CV}}_{\mathrm{PQS1}}$ in effect are functions of $\delta$, $t_0$, $\varphi$, and $t$.

As for the PQS2 proposed in \cref{subsec:SecondImplementation}, applying it  on mode 2 of the state $\ket{\Xi}_{12}$ yields the (unnormalized) output (see Eqs.\ \eqref{eq:OutputPQS2PsiP} to \eqref{eq:FidePQS2PsiP})
\begin{eqnarray}
&& K_1 (-i\Gamma) N_0 f_1(\beta) ( \ket{1_H}_2 \ket{\alpha_H}_1 \!-\! e^{i\varphi} \ket{1_V}_2 \ket{\! - \alpha_V}_1 ) \nonumber \\
 && + \! K_0 (-i \Gamma)^2 N_0 L_\alpha^{-1} f_0 (\beta) \ket{1_H,\!1_V}_2 L_\alpha (\ket{\alpha_H} \!+\! e^{i\varphi} \ket{\!-\alpha_V})_1. \nonumber \\ \label{eq:OutputHybridPQS2}
\end{eqnarray}
The success probability and the fidelity of the entanglement preparation for this case are
\begin{eqnarray}
 P^{\mathrm{DV\text{-}CV}}_{\mathrm{PQS2}} &=& 2 K_1^2 |\Gamma|^2 N_0^2 f_1 (\beta)^2  + K_0^2 |\Gamma|^4 N_0^2 L_\alpha^{-2} f_0^2(\beta),  \nonumber \\ \label{eq:ProbHy2} \\
 F^{\mathrm{DV\text{-}CV}}_{\mathrm{PQS2}} &=& \frac{2 K_1^2 f_1 (\beta)^2 }{2 K_1^2 f_1 (\beta)^2   + K_0^2 |\Gamma|^2 L_\alpha^{-2} f_0^2(\beta)},  \label{eq:FideHy2}
\end{eqnarray}
which are actually dependent on $\delta $, $t_0$, $\varphi$, and $\Gamma$. Similar to the case of the PQS1, the PQS2 also produces an undesired term (i.e., the term on the second line of Eq.\ \eqref{eq:OutputHybridPQS2}) of which mode 2 is a two-photon state. This term gives rise to an increase in the heralding probability [due to the second term in the RHS of Eq.\ \eqref{eq:ProbHy2}] at the cost of decreasing the corresponding fidelity [due to the second term in the denominator of the RHS of Eq.\ \eqref{eq:FideHy2})].

\begin{figure}[t!]
    \centering
    \includegraphics[scale=0.62]{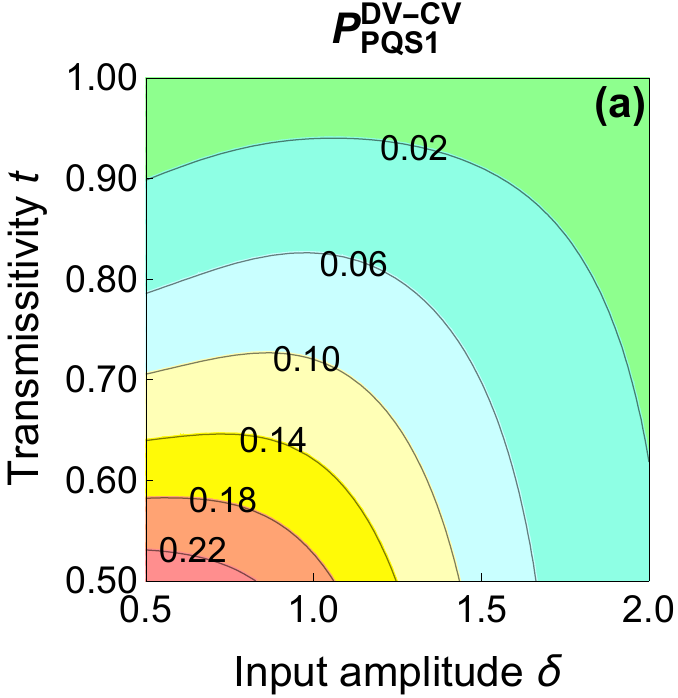}\hfill
    \includegraphics[scale=0.62]{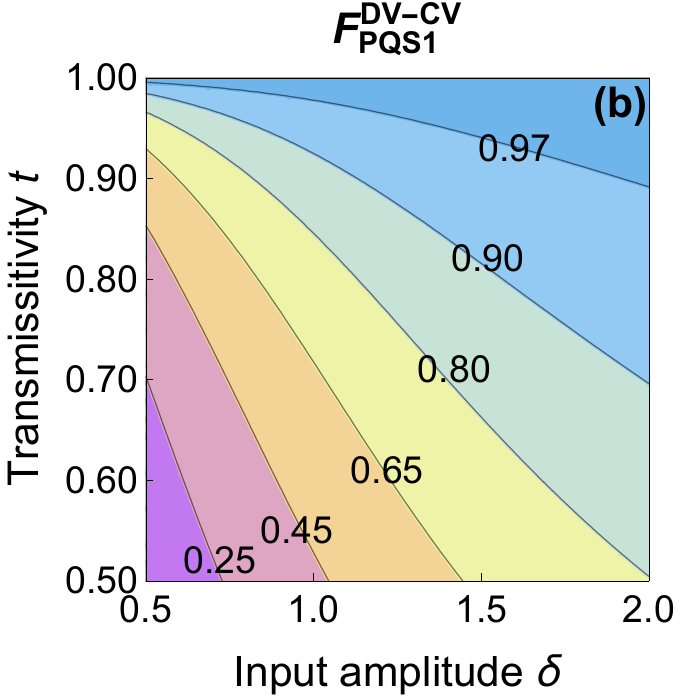}\vspace{0.2cm}
     \includegraphics[scale=0.62]{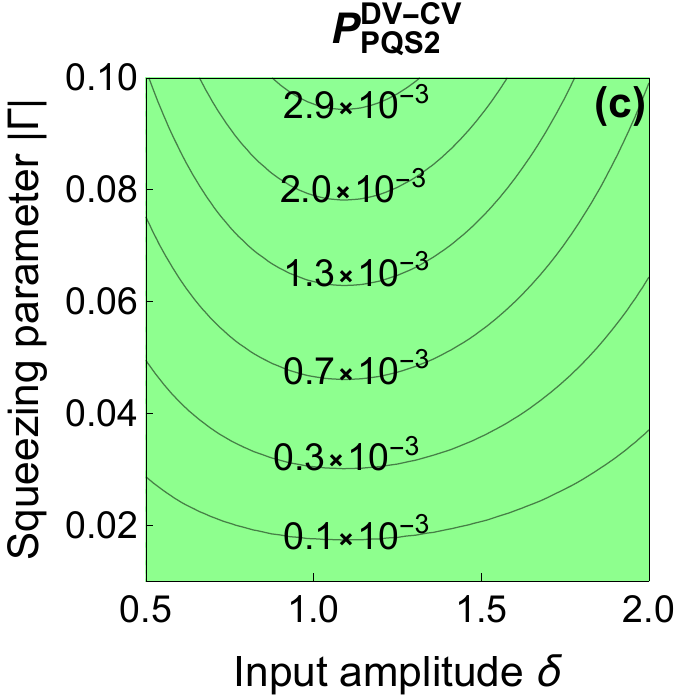}\hfill
    \includegraphics[scale=0.62]{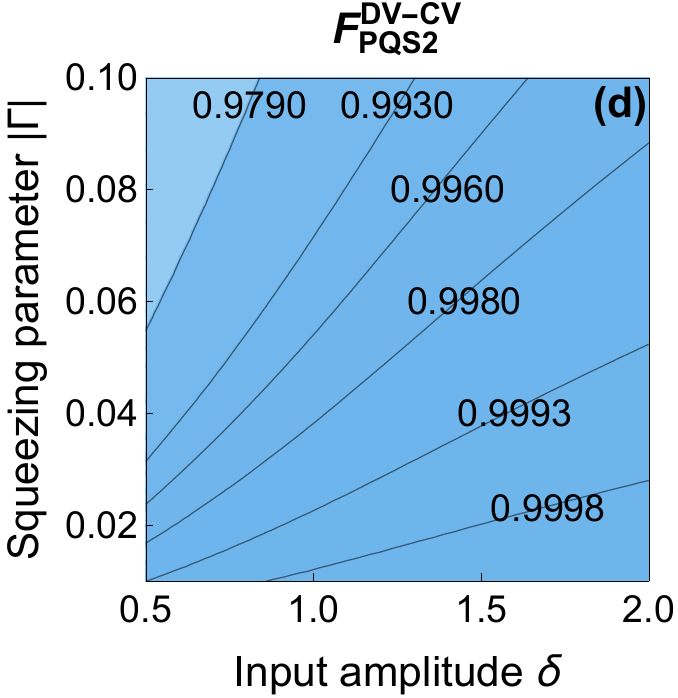} 
    \caption{(a), (b) Success probability $P^{\mathrm{DV\text{-}CV}}_{\mathrm{PQS1}}$ and fidelity $F^{\mathrm{DV\text{-}CV}}_{\mathrm{PQS1}}$ given in Eqs.\ \eqref{eq:ProbHy1} and \eqref{eq:FideHy1}, respectively, as functions of the input amplitude $\delta$ and the transmissivity $t$ for the preparation of the hybrid entanglement $\ket{\Phi^{\mathrm{DV\text{-}CV}}}$ in \cref{eq:DVCV} using the PQS1 developed in \cref{subsec:FirstImplementation}. (c), (d)  Success probability $P^{\mathrm{DV\text{-}CV}}_{\mathrm{PQS2}}$ and fidelity $F^{\mathrm{DV\text{-}CV}}_{\mathrm{PQS2}}$ given in Eqs.\ \eqref{eq:ProbHy2} and \eqref{eq:FideHy2}, respectively, as functions of the input amplitude $\delta$ and the squeezing parameter $|\Gamma|$ for the preparation of the same entanglement but using the PQS2 developed in \cref{subsec:SecondImplementation}. In plotting these figures  $\phi = 0$ and $t_0 = 0.5$ were chosen.  }
    \label{fig:Hybrid}
\end{figure}

To display our results graphically we choose $\varphi=0$ and $t_0=0.5$, which give $\alpha = \beta =\delta$ and $\ket{\Xi}_{12} = N_0(\ket{\delta_H}_1\ket{\delta_H}_2 + \ket{\!-\delta_V}_1 \ket{-\delta_V}_2)$). With such parameters we plot $P^{\mathrm{DV\text{-}CV}}_{\mathrm{PQS1}}$ and $F^{\mathrm{DV\text{-}CV}}_{\mathrm{PQS1}}$ as functions of the input amplitude $\delta$ and the transmissitivity $t$ in Figs.\ \ref{fig:Hybrid}a and \ref{fig:Hybrid}b and  $P^{\mathrm{DV\text{-}CV}}_{\mathrm{PQS2}}$ and $F^{\mathrm{DV\text{-}CV}}_{\mathrm{PQS2}}$ as functions of the input amplitude  $\delta$ and the squeezing parameter $|\Gamma|$ in Figs.\ \ref{fig:Hybrid}c and \ref{fig:Hybrid}d. For the parameter ranges in \cref{fig:Hybrid}, we observe that the PQS1 is superior to the PQS2 in terms of the success probability but is inferior in terms of the fidelity. Discretely, $P^{\mathrm{DV\text{-}CV}}_{\mathrm{PQS1}}$ is of order $10^{-2} - 10^{-1}$, much larger than $P^{\mathrm{DV\text{-}CV}}_{\mathrm{PQS2}}$, which is in the range $10^{-4} - 10^{-3}$.  $F^{\mathrm{DV\text{-}CV}}_{\mathrm{PQS1}}$ varies quite largely from about 0.25 to above 0.97, whereas $F^{\mathrm{DV\text{-}CV}}_{\mathrm{PQS2}}$ remains very close to 1 for almost the whole domain of the used parameters. These manifest the pros and cons of the two PQS implementations.   

Closer looking at $P^{\mathrm{DV\text{-}CV}}_{\mathrm{PQS1}}$ in \cref{fig:Hybrid}a and $F^{\mathrm{DV\text{-}CV}}_{\mathrm{PQS1}}$ in \cref{fig:Hybrid}b reveals two apparently contrast patterns in their variations with respect to $\delta$ and $t$: $P^{\mathrm{DV\text{-}CV}}_{\mathrm{PQS1}}$ decreases when increasing both $\delta$ and $t$ while $F^{\mathrm{DV\text{-}CV}}_{\mathrm{PQS1}}$ improves. This is understandable, since from Eqs.\ \eqref{eq:ProbHy1} and \eqref{eq:FideHy1} we find that $\lim_{t\to1} P^{\mathrm{DV\text{-}CV}}_{PQS1} = 0$ and $\lim_{t\to 1}F^{\mathrm{DV\text{-}CV}}_{PQS1} = 1$, indicating that the state-preparation fidelity can be made arbitrarily high by adjusting the transmissitivity $t$ to close to 1 but with a price of an unrealistically low heralding probability. We also note that increasing $\delta$, roughly speaking, leads to a decrease in $f_0^2(\beta) \equiv f_0^2(\delta)$ and   $f_1^2(\beta) \equiv f_1^2(\delta)$ and an increase in $f_1^2(\beta)/f_0^2(\beta) \equiv \delta^2$. The former directly links to a reduction of $P^{\mathrm{DV\text{-}CV}}_{\mathrm{PQS1}}$ (see Eq.\ \eqref{eq:ProbHy1}); the latter is equivalent to lessening the contribution of the unwanted term in the output state in Eq.\ \eqref{eq:OutputHybridPQS1} and thus enhancing $F^{\mathrm{DV\text{-}CV}}_{\mathrm{PQS1}}$. 
As for  $P^{\mathrm{DV\text{-}CV}}_{\mathrm{PQS2}}$ and $F^{\mathrm{DV\text{-}CV}}_{\mathrm{PQS2}}$, the first quantity increases with $|\Gamma|$ and for a given $|\Gamma|$ is maximized by a particular value of $\delta$ which is numerically found to be near 1. The second quantity changes very slowly with $\delta$ and  $|\Gamma|$ and stays close to 1, which is not surprising if we look at its expression in Eq.\ \eqref{eq:FideHy2}. For $|\Gamma| = 10^{-2} - 10^{-1} \ll 1$  the second term in the denominator of $F^{\mathrm{DV\text{-}CV}}_{\mathrm{PQS2}}$ is negligible, making
$F^{\mathrm{DV\text{-}CV}}_{\mathrm{PQS2}} \simeq 1$.

\subsection{DV-DV polarization Bell pair}

\begin{figure}[ht!]
    \centering
    \includegraphics[scale=0.62]{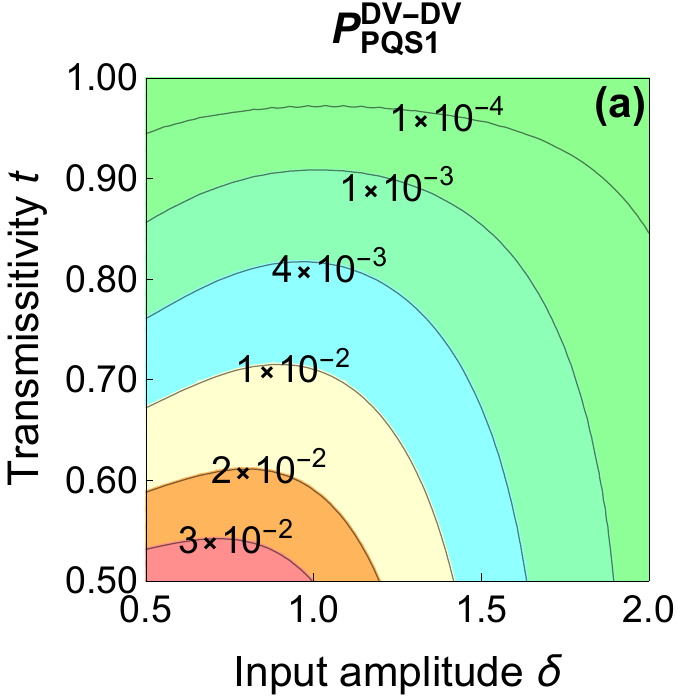}\hfill
    \includegraphics[scale=0.62]{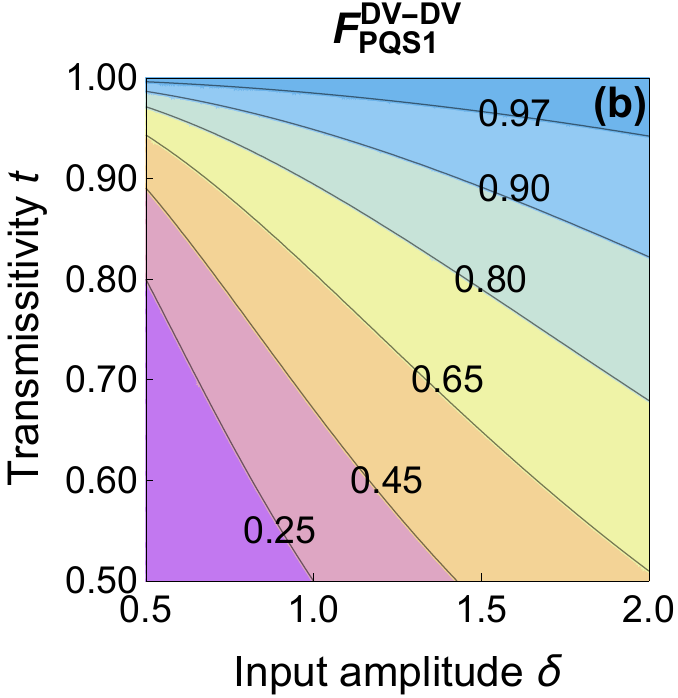}\vspace{0.2cm}
     \includegraphics[scale=0.62]{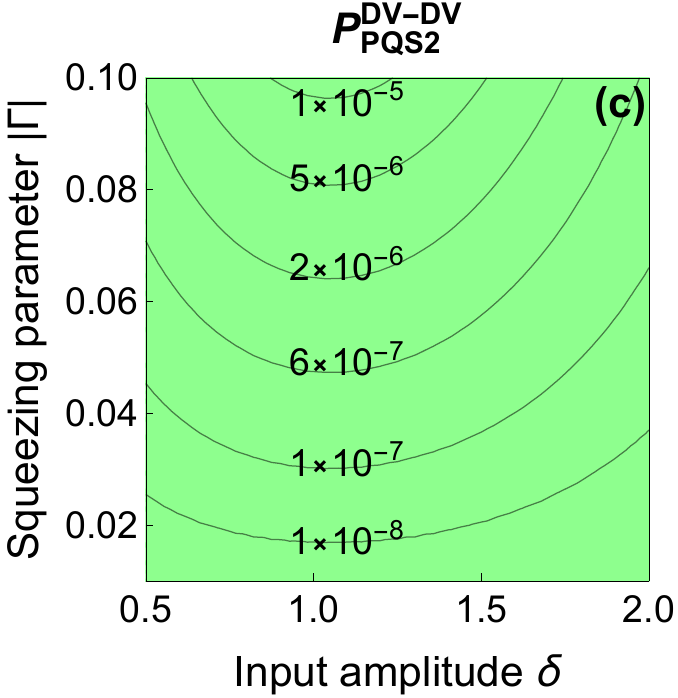}\hfill
    \includegraphics[scale=0.62]{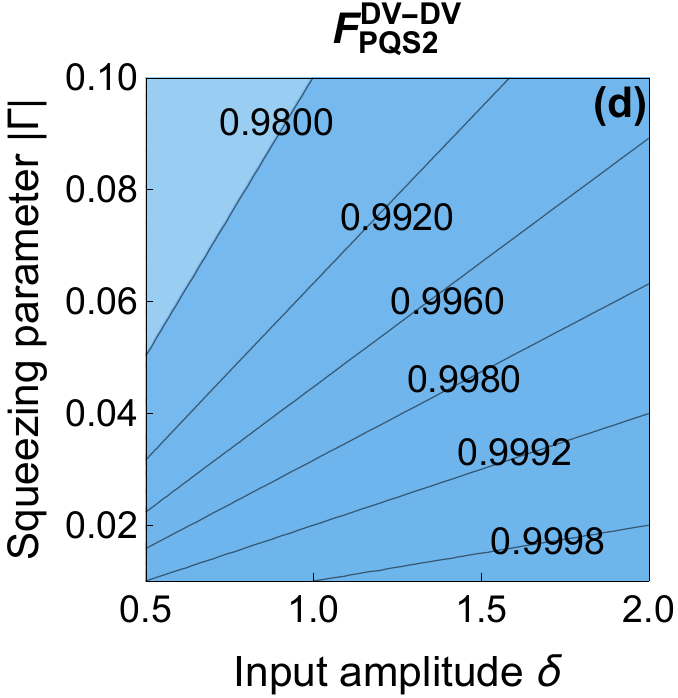} 
    \caption{(a), (b) Success probability $P^{\mathrm{DV\text{-}DV}}_{\mathrm{PQS1}}$ and fidelity $F^{\mathrm{DV\text{-}DV}}_{\mathrm{PQS1}}$ given in \cref{append:BellPair} as functions of the input amplitude $\delta$ and the transmissivity $t$ for the preparation of the PBP $\ket{\Phi^{\mathrm{DV\text{-}DV}}}$ in \cref{eq:DVDV} using the PQS1 developed in \cref{subsec:FirstImplementation}. (c), (d)  Success probability $P^{\mathrm{DV\text{-}DV}}_{\mathrm{PQS2}}$ and fidelity $F^{\mathrm{DV\text{-}DV}}_{\mathrm{PQS2}}$ given in \cref{append:BellPair} as functions of the input amplitude $\delta$ and the squeezing parameter $|\Gamma|$ for the preparation of the same entanglement but using the PQS2 developed in \cref{subsec:SecondImplementation}. In plotting these figures  $\phi = 0$ and $t_0 = 0.5$ were chosen.  }
    \label{fig:BellPair}
\end{figure}
Given the output state in Eq.\ \eqref{eq:OutputHybridPQS1} or \eqref{eq:OutputHybridPQS2}, that is close to the hybrid entangled state $( \ket{1_H}_2 \ket{\alpha_H}_1 - e^{i\varphi} \ket{1_V}_2 \ket{\!-\alpha_V}_1 )/\sqrt{2}$, we continue truncating irrelevant components in mode 1 to get the PBP  $\ket{\Phi^{\mathrm{DV\text{-}DV}}}$ in \cref{eq:DVDV} via the PQS1 as well as the PQS2. Detailed calculations for such processes are provided in \cref{append:BellPair} and here we highlight the main results only.
Figure \ref{fig:BellPair} shows the performance for the preparation of the PBP using the two PQSs. The patterns for the success probabilities and the fidelities in \cref{fig:BellPair} are substantially similar to those in \cref{fig:Hybrid}, except the success probabilities are now lower by several orders due to double truncation.
We again see an evident trade-off in the performances of the two QPS implementations. The QPS1 yields a good heralding probability but a modest fidelity, while the QPS2 features at a very low success probability and a close-to-unit fidelity.

\section{Discussion} \label{sec:Discussion}

\begin{figure}[ht!]
    \centering
    \includegraphics[width = 0.48 \textwidth]{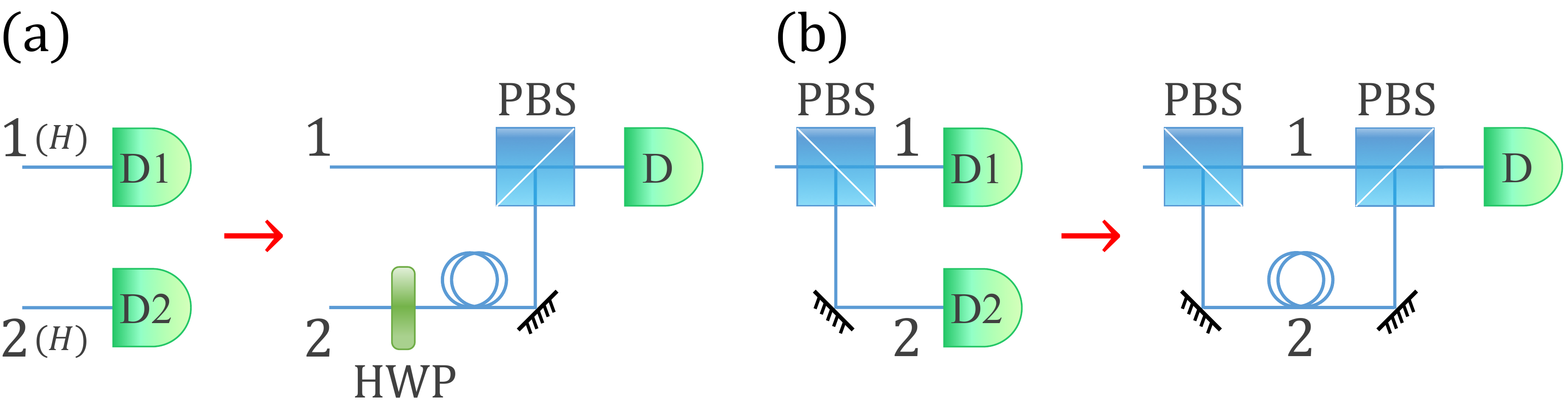}
    \caption{(a) Reduction of the number of  photodetectors in the PQS1 implementation. In the left, two spatially separated photons in modes $1$ and $2$ of the same horizontal $(H)$ polarization are detected by two detectors D1 and D2. In the right, employing a HWP, a delay circle, a mirror, and a PBS makes the two photons detected by only one detector D but at adjustably separated arrival times. (b) Similar to (a) but for the PQS2 implementation with two detected photons initially having orthogonal polarizations.} 
    
    \label{fig:ReducePNRDetectors}
\end{figure}

The results in the \cref{sec:Performance} clearly show the advantages and disadvantages  of each of the two PQS implementations in the entanglement production.  The PQS1 is beneficial when it comes to the heralding probability but possesses a moderate fidelity. The PQS2, in contrast, performs with a low success probability but  yields a very high fidelity.  Concerning the consumed resources, the PQS1 requires two ancilla photons of orthogonal polarizations and four photon-number-resolving (PNR) detectors. The PQS2  differently involves one SPDC crystal accompanied by optical pumping to stimulate squeezing and two PNR detectors. 
Notably, owing to the interplay between polarization and spatial DoFs \cite{KokBook10} we can reduce the number of PNR detectors in both the PQS implementations. Concretely, instead of detecting two spatially separated photons by two detectors as in Figs.\ \ref{fig:QuantumScissors} and \ref{fig:DatQuantumScissors} we exploit delay circles, mirrors, HWPs, and PBSs, as illustrated in \cref{fig:ReducePNRDetectors}, to detect the  photons by only one detector at the same spatial mode but at different arrival times. By this the number of needed PNR detectors in the PQS1 implementation is decreased from 4 to 2, whereas that in the PQS2 implementation becomes 1.

The heralded entanglement preparation proposed in this paper crucially depends on the availability of the input CV entanglements in Eqs.\  \eqref{eq:CVEntanglement2modes} and \eqref{eq:CVEntanglementnmodes}. These entangled states as demonstrated in \cref{fig:CVEntanglement2modes} are produced from a coherent state and a cat state. The latter is of great importance in foundations of the quantum theory \cite{Schrodinger35} and quantum applications \cite{Ralph03,Lund08} but in general  is troublesome to prepare \cite{Mikheev19}. Thankfully, in practice there are states that can be used as cat states with high fidelities. Namely,  kitten  states, i.e., cat states with small amplitudes (up to 1), can be very well approximated by a deterministic squeezed vacuum \cite{Israel19}. Large-amplitude cat states (larger than $2$) can be prepared probabilistically via breeding of kitten states \cite{Sychev17catstate} or generalized photon subtraction \cite{Takase21}. Moreover, single-photon inputs required in the PQS1 can be supplied deterministically with high quality by semiconductor quantum-dot emitters \cite{Senellart17,Tomm21,Li21}.  
Another salient ingredient in our entanglement generation scheme as well as in many other quantum optical protocols is PNR detectors.
Superconducting transition edge sensors PNR detectors \cite{Lita08} have recently been shown to operate with an efficiency exceeding $95 \%$ \cite{Thekkadath20,Arrazola21}. Another possibilities to discriminate photon numbers include multiplexing of single-photon detectors \cite{Israel19,Jonsson19,Provaznik20,Nehra20} and fine analyses of output signal waveforms \cite{Endo21}. We note that for a small-amplitude input state (i.e., the state $\ket{\Xi}_{12}$ in \cref{eq:CVEntanglement2modes} with small $\alpha$ and $\beta$) of which more-than-one-photon contributions are small compared to those of the vacuum and the single-photon state, non-PNR detectors such as single-photon counting modules (SPCMs) \cite{Wagenknecht10} or single-photon avalanche photodiodes (SPADs) \cite{Nehra20} might suffice for our scheme.

We compare the entanglement preparation scheme proposed here with the existing ones in the literature \cite{Kwon15,Li18,Gouzien20,Le21,Barz10,Wagenknecht10,Sliwa03,Li21}.  Our scheme to generate the DV-CV hybrid entangled state in Eq.\ \eqref{eq:DVCV}, in comparisons with those in Refs.\ \cite{Kwon15,Li18,Gouzien20,Le21}, does not require a PBP input and uses a lower number of PNR detectors, thus relaxing initial overheads. Also, the success probability and the prepared-state fidelity of our scheme at a proper choice of relevant parameters can be made considerably higher than those in Refs.\ \cite{Kwon15,Li18,Gouzien20,Le21}.  
As for the DV-DV PBP preparation, we estimate the count rate in our scheme and compare it to those in Refs.\ \cite{Barz10,Wagenknecht10,Li21}. In particular,  we assume  that the input entangled state $\ket{\Xi}_{12}$ prepared deterministically from a squeezed vacuum and the input single photons emitted from a quantum-dot emitter are supplied on-demand with repetition rate  $6.4\,\mathrm{MHz}$ \cite{Li21}. We then employ the PQS1, choose $\delta = 0.8$ and $t = 0.98$ at which $P^{\mathrm{DV\text{-}DV}}_{\mathrm{PQS1}} \sim 3.6 \times 10^{-5}$ and $F^{\mathrm{DV\text{-}DV}}_{\mathrm{PQS1}} > 0.9 $ (see Figs.\ \ref{fig:BellPair}a and \ref{fig:BellPair}b), and adapt multiplexing PNR detectors in Refs.\ \cite{Nehra20,Israel19} that are compatible with MHz repetition rates to generate the PBP at a count rate approximately given by $6.4\,\mathrm{MHz}\times 3.6\times 10^{-5} \approx 230\, \mathrm{Hz}$.  
Using the PQS2, we operate it at $\delta = 0.8$ and $|\Gamma| = 0.07 $ with $P^{\mathrm{DV\text{-}DV}}_{\mathrm{PQS2}} \sim 2\times 10^{-6} $ and $F^{\mathrm{DV\text{-}DV}}_{\mathrm{PQS2}} > 0.98 $ (see Figs.\ \ref{fig:BellPair}c and \ref{fig:BellPair}d) and again assume that the entangled input $\ket{\Xi}_{12}$ is supplied from a squeezed vacuum but with a higher repetition rate of $80\,\mathrm{MHz}$ \cite{Israel19}, so that the PBP preparation  will  have a count rate roughly at $ 80\,\mathrm{MHz} \times 2\times 10^{-6}  = 160\, \mathrm{Hz}$. The count rates estimated above for preparing the PBP with fidelities $> 0.9$ are  several orders higher than those reported in Refs.\  \cite{Barz10,Wagenknecht10,Li21}. Additionally, the use of a relatively small squeezing parameter $|\Gamma|$ in the PQS2 here eases up the requirement for very strong pumping and reduces the effect of higher  squeezing orders which can lead to false detections as in Refs.\ \cite{Barz10,Wagenknecht10}.      

\section{Conclusion} \label{sec:Conclusion}

We presented a general scheme to prepare $n$-partite polarization entangled states for an arbitrary $n\ge 2$ and considered two specific examples, namely, the hybrid DV-CV entangled state between polarized single photons and polarized coherent states  in \cref{eq:DVCV} and the DV PBP between polarized single photons in \cref{eq:DVDV}.
Our scheme involves neither postselection nor destruction of photons\cite{Kwiat95,Kwiat98}. Different from the existing heralded schemes \cite{Zou02,Zou02Third,Pittman03,Kwon15,Li18,Gouzien20,Le21,Barz10,Wagenknecht10,Sliwa03,Li21}, we harness the connections among CV, hybrid DV-CV, and DV entanglements and propose a truncation technique, i.e., the PQS, to map a given CV entanglement into the desired entangled states in a heralded fashion.
The needed input state is an entangled coherent state in \cref{eq:CVEntanglement2modes} or Eq.\ \eqref{eq:CVEntanglementnmodes} which can be supplied by modern quantum technologies following the schematic setup in \cref{fig:CVEntanglement2modes}.  We designed two different PQS implementations, the PQS1 in \cref{fig:QuantumScissors} and the PQS2 in \cref{fig:DatQuantumScissors}. The PQS1 employs a single-photon source and linear-optics devices while the PQS2 uses a two-mode squeezer which is a commonplace optical nonlinear 
element. The former is advantageous in success probability and the later excels in fidelity, as shown by detailed performance analysis of the entanglement preparation in \cref{sec:Performance}. 
Both the PQS1 and the PQS2 could be implemented within the current optical toolbox. Effects of imperfect input states, inefficient PNR detectors, nonideal device operations, and decoherence from surrounding environments are beyond the scope of the present paper but will be the subject of a subsequent work. Also, applications of the PQSs presented here to other quantum information tasks such as noiseless linear amplification \cite{Winnel20,He21} and studies of the transition from CV entanglement to DV entanglement in the context of the nonclassicality and classical simultability \cite{Andersen15,Rundle20,Alvarez20}  are worth pursuing.

\begin{acknowledgments}
In this work D.T.L. is supported by the Australian Research Council Centre of Excellence for Engineered Quantum Systems (EQUS, CE170100009),
and N.B.A. is supported by the National Foundation for Science and Technology Development (NAFOSTED) under Project No. 103.01-2019.313.
\end{acknowledgments}

\appendix

\section{ Success probability and fidelity for the preparation of the polarization Bell pair} \label{append:BellPair}

\subsection{ Using the PQS1}

We rewrite the output in Eq.\ \eqref{eq:OutputHybridPQS1} as
\begin{equation}
    \ket{\alpha_H}_1 (g_1 \ket{1_H} + g_0 \ket{0} )_2 + \ket{\! - \alpha_V}_1 (-g_1 e^{i\varphi} \ket{1_V} +g_0 e^{i\varphi} \ket{0})_2,
\end{equation}
where 
\begin{equation}
    g_1 = \sqrt{(1-t)t} N_0 f_1 (\beta), \hspace{0.5cm} g_0 = (1-t)N_0 f_0 (\beta).
\end{equation}
We use the PQS1 in \cref{subsec:FirstImplementation} to truncate non-single-photon components in mode $1$ of this state and obtain the output
\begin{eqnarray}
 && \sqrt{(1\!-\!t)t} f_1 (\alpha) \ket{1_H}_1 (g_1 \ket{1_H} + g_0 \ket{0} )_2 \nonumber \\
 &+&  \sqrt{(1\!-\!t)t} f_1 (-\alpha) \ket{1_V}_1 (-g_1 e^{i\varphi} \ket{1_V} +g_0 e^{i\varphi} \ket{0})_2 \nonumber \\
 &+&  (1\!-\!t) f_0 (\alpha) \ket{0}_1 \big[ g_1(\ket{1_H} - e^{i\varphi} \ket{1_V} ) + g_0 (1+e^{i\varphi}) \ket{0} \big]_2. \nonumber \\
\end{eqnarray}
The success probability and the fidelity of such output compared to the desired PBP $\ket{\Phi^{\mathrm{DV\text{-}DV}}}$ in \cref{eq:DVDV} are
\begin{eqnarray}
 P^{\mathrm{DV\text{-}DV}}_{\mathrm{PQS1}} \! &=& \! \big[2 (1-t)t f_1^2(\alpha) + 2 (1-t)^2 f_0^2(\alpha) \big] (g_1^2 + g_0^2),  \nonumber \\
 \\
 F^{\mathrm{DV\text{-}DV}}_{\mathrm{PQS1}} \! &=& \! \frac{ t f_1^2(\alpha) g_1^2}{\big[ t f_1^2(\alpha) +  (1-t) f_0^2(\alpha) \big] (g_1^2 + g_0^2)}. 
\end{eqnarray}

\subsection{ Using the PQS2}

We rewrite the output in Eq.\ \eqref{eq:OutputHybridPQS2} as
\begin{equation}
    \ket{\alpha_H}_1 (h_1 \ket{1_H} + h_0 \ket{0} )_2 + \ket{\! - \alpha_V}_1 (-h_1 e^{i\varphi} \ket{1_V} +h_0 e^{i\varphi} \ket{0})_2,
\end{equation}
where
\begin{equation}
    h_1 = K_1 (-i\Gamma) N_0 f_1 (\beta), \hspace{0.2cm}  h_0 = K_0 (-i \Gamma)^2 N_0 f_0 (\beta). 
\end{equation}
We use the PQS2 in \cref{subsec:SecondImplementation} to truncate non-single-photon components in mode $1$ of this state and obtain the output
\begin{eqnarray}
 && K_1(-i\Gamma) f_1 (\alpha) \ket{1_H}_1 (h_1 \ket{1_H} + h_0 \ket{0} )_2 \nonumber \\
 &+& K_1(-i\Gamma) f_1 (- \alpha) \ket{1_V}_1 (-h_1 e^{i\varphi} \ket{1_V} +h_0 e^{i\varphi} \ket{0})_2 \nonumber \\
 &+&  K_0 (-i \Gamma)^2 f_0 (\alpha) \ket{1_H,1_V}_1 \nonumber \\
 &\times &  \big[ h_1(\ket{1_H} - e^{i\varphi} \ket{1_V} ) + h_0 (1+e^{i\varphi}) \ket{0} \big]_2.
\end{eqnarray}
The success probability and the fidelity of such output compared to the desired PBP $\ket{\Phi^{\mathrm{DV\text{-}DV}}}$ in \cref{eq:DVDV} are
\begin{eqnarray}
 P^{\mathrm{DV\text{-}DV}}_{\mathrm{PQS2}} \! &=& \! \big[ 2K_1^2 |\Gamma|^2 f_1^2 (\alpha) \! + \! 2K_0^2 |\Gamma|^4 f_0^2 (\alpha) \big] (|h_1|^2 \! + \! |h_0|^2),  \nonumber \\ \\
 F^{\mathrm{DV\text{-}DV}}_{\mathrm{PQS2}} \! &=& \!  \frac{ K_1^2  f_1^2 (\alpha) |h_1|^2 }{\big[ K_1^2  f_1^2 (\alpha) + K_0^2 |\Gamma|^2 f_0^2 (\alpha) \big] (|h_1|^2 + |h_0|^2)}. \nonumber \\
\end{eqnarray}

\bibliography{heraldGen}

\end{document}